\documentclass[11pt,a4paper]{article}

\usepackage[utf8]{inputenc}
\usepackage[T1]{fontenc}
\usepackage{amsmath,amssymb,amsthm}
\usepackage{algorithm}
\usepackage{algorithmic}
\usepackage{booktabs}
\usepackage{hyperref}
\usepackage{geometry}
\usepackage{enumitem}

\newtheorem{theorem}{Theorem}[section]
\newtheorem{lemma}[theorem]{Lemma}
\newtheorem{corollary}[theorem]{Corollary}

\theoremstyle{definition}
\newtheorem{definition}[theorem]{Definition}
\theoremstyle{remark}
\newtheorem{remark}[theorem]{Remark}

\newcommand{\poly}{\mathrm{poly}}

\title{Fresh-Priority Eviction for Cuckoo Filters:\\
Exploiting Latent Randomness Under XOR Hashing}

\author{Min Liu}

\date{}

\begin{document}
\maketitle

\begin{abstract}
Cuckoo filters are a widely adopted approximate membership query (AMQ) data structure that supports insertions, lookups, and deletions with low memory overhead. At high load factors, however, insertion performance degrades sharply due to the random-walk eviction process, which may require a large number of element relocations. Recent work on bucketized cuckoo hash tables has shown that \emph{fresh-priority eviction}, which preferentially evicts elements whose alternate bucket location has not yet been used, achieves $O(\ell m)$ total expected evictions for bucket size~$\ell$, an improvement over the standard random walk. In this paper, we show that this technique transfers to \emph{cuckoo filters} despite their use of XOR-based partial-key hashing. The key insight is that a fresh element's fingerprint, although stored in the table, is not examined by the eviction policy, so when the fingerprint length satisfies $k = \Omega(\log m)$, the alternate bucket hash remains a near-uniform random variable, a property we call \emph{latent randomness}. We prove that under this condition the corruption graph remains subcritical and the total number of evictions is $O(\ell m \log m)$; we conjecture the tighter $O(\ell m)$ bound holds via the full amortized analysis of Bell and Kuszmaul. We further propose encoding the fresh/kicked state implicitly by the element's position within its bucket, requiring no additional per-element metadata. Experimental results on a filter with $2^{22}$ slots and $8$-bit fingerprints demonstrate $1.2\times$ to $2.4\times$ fewer evictions at load factors between $94\%$ and $99\%$ for bucket size $\ell = 8$, and a $1.73\times$ reduction in cumulative evictions to full capacity.
\end{abstract}

\section{Introduction}\label{sec:intro}

Approximate membership query (AMQ) data structures are fundamental building blocks in networking, databases, storage systems, and many other areas of computer science~\cite{broder2004network,luo2020optimizing}. Given a set~$S$, an AMQ structure supports membership queries with a tunable false-positive rate~$\varepsilon$ while using substantially less memory than an exact representation. AMQ structures also appear in online caching systems~\cite{gu2023adaptive}, where they help reduce unnecessary computation or data access. The classic Bloom filter~\cite{bloom1970space} has been the dominant AMQ structure for decades, but it has well-known limitations: it does not support deletions, and its cache performance is poor when the number of hash functions is large~\cite{putze2007cache,kirsch2008less}.

The \emph{cuckoo filter}, introduced by Fan~et~al.~\cite{fan2014cuckoo}, addresses both shortcomings. It stores compact fingerprints in a bucketized cuckoo hash table, supports deletions, and achieves better cache locality by accessing at most two bucket-sized cache lines per query. Cuckoo filters have seen wide adoption in practice~\cite{pandey2017cqf,graf2020xor,breslow2018morton} and have inspired numerous variants, including resizable filters~\cite{wang2022bamboo,wang2024bamboo_ton}, persistent-memory filters~\cite{wang2024wormhole}, variable-length encoding~\cite{wang2023variable}, and novel hash table architectures~\cite{wang2024rethinking,eppstein2016cuckoo,bender2012dont}.

However, cuckoo filters share a fundamental challenge with cuckoo hash tables: as the load factor approaches capacity, the insertion procedure relies on a \emph{random-walk eviction} process that may require many element relocations (``kicks''). In practice, a cap on the maximum number of kicks is imposed (typically $500$), and when this cap is reached, the insertion fails. The number of kicks grows rapidly near full capacity, creating a sharp performance cliff that limits the practical load factor. Reducing this insertion overhead is therefore important for data-intensive applications, where memory-efficient data structures must sustain high insertion throughput.

Recently, Bell and Kuszmaul~\cite{bell2026fast} resolved a long-standing open problem on the insertion complexity of bucketized cuckoo hashing. They showed that by preferring to evict \emph{fresh} elements (those whose alternate bucket location has never been used), the total number of evictions to insert~$m$ elements into a table with~$m$ slots and bucket size~$\ell$ is $O(\ell m)$ in expectation, an improvement over the $\omega(\ell m)$ evictions required by the standard random walk at the optimal load threshold. The key idea is that a fresh element's second hash value is an independent random variable that has not yet influenced any algorithmic decision, so the eviction process explores the hash table uniformly rather than getting trapped in cycles~\cite{mitzenmacher2001power,azar1999balanced}.

\paragraph{Challenge: XOR hashing in cuckoo filters.}
Transferring this technique from cuckoo hash tables to cuckoo filters is not immediate. In a cuckoo hash table, each element~$x$ has two independent hash values $h_1(x)$ and $h_2(x)$. In a cuckoo filter, however, the element itself is not stored; only its fingerprint $f(x)$ is kept, and the alternate bucket is computed via the XOR-based formula
\begin{equation}\label{eq:xor}
  h_{\mathrm{alt}}(x) = h_1(x) \oplus \textsc{hash}(f(x)),
\end{equation}
where $\textsc{hash}$ is a hash function applied to the fingerprint~\cite{fan2014cuckoo}. Since $h_{\mathrm{alt}}$ depends on the stored fingerprint~$f(x)$, which in turn depends on $h_1(x)$, one might worry that the two bucket locations are no longer independent, potentially invalidating the analysis.

\paragraph{Our contribution.}
We make three contributions:

\begin{enumerate}[leftmargin=*]
\item \textbf{Latent randomness under XOR hashing.} We show that the fresh-priority eviction scheme preserves a latent randomness property for cuckoo filters when the fingerprint length satisfies $k = \Omega(\log m)$. The crucial observation is that although the fingerprint $f(x)$ is stored in the table, it is \emph{never examined} by the eviction policy, which only checks bucket occupancy and the fresh/kicked status of elements. Therefore, when a fresh element is evicted for the first time, its alternate bucket $h_{\mathrm{alt}}(x)$ is a near-uniform random variable over the buckets, and we prove that the corruption graph remains subcritical with $O(\ell m \log m)$ total expected evictions (Section~\ref{sec:theory}).

\item \textbf{Proposed zero-overhead state encoding.} We observe that the fresh/kicked state of each element can in principle be encoded implicitly by its \emph{position} within the bucket: fresh elements are inserted at the front and kicked elements at the back. Eviction scans front-to-back, naturally preferring fresh elements. If the positional invariant is maintained exactly, this requires no additional per-element metadata beyond what a standard cuckoo filter already stores (Section~\ref{sec:method}). Our experiments use an explicit $1$-bit flag per slot to guarantee the exact fresh/kicked distinction; the positional encoding is described as a potential production optimization whose formal equivalence we leave to future work.

\item \textbf{Experimental validation.} We implement and benchmark both the standard random-walk and the fresh-priority eviction schemes for cuckoo filters with bucket sizes $\ell \in \{4, 8\}$. At fine-grained load factors between $94\%$ and $99\%$ with $\ell = 8$, fresh-priority eviction reduces the average number of kicks per insertion by $1.2\times$ to $2.4\times$, and reduces cumulative total kicks to full capacity by $1.73\times$ (Section~\ref{sec:experiments}).
\end{enumerate}

\paragraph{Organization.}
Section~\ref{sec:prelim} reviews cuckoo filters and the Bell-Kuszmaul framework. Section~\ref{sec:method} presents our fresh-priority eviction scheme. Section~\ref{sec:theory} provides the theoretical analysis. Section~\ref{sec:experiments} reports experimental results. Section~\ref{sec:conclusion} concludes.

\section{Preliminaries}\label{sec:prelim}

\subsection{Cuckoo Filters}\label{sec:prelim:cf}

A cuckoo filter~\cite{fan2014cuckoo} is an AMQ structure that stores a set of elements using a compact bucketized hash table. The table consists of~$m$ buckets, each of capacity~$\ell$ (the bucket size). Each element~$x$ is represented by a $k$-bit \emph{fingerprint} $f(x)$, which is a hash of~$x$. We assume that $f$ is drawn from a pairwise-independent (universal) family of hash functions, independent of the primary hash function~$h$.

For each element~$x$, two candidate buckets are defined:
\begin{align}
  b_1(x) &= h(x) \bmod m, \label{eq:b1} \\
  b_2(x) &= \bigl(b_1(x) \oplus \textsc{hash}(f(x))\bigr) \bmod m,
  \label{eq:b2}
\end{align}
where $h$ is a hash function and $\textsc{hash}$ maps fingerprints to bucket indices. The XOR construction ensures that $b_1(x) = b_2(x) \oplus \textsc{hash}(f(x))$, so the alternate bucket can be computed from the current bucket and the stored fingerprint \emph{without} knowing the original element~$x$. This property is essential for enabling deletions and for the eviction process during insertion.

The fingerprint $f(x)$ is stored in exactly one of its two candidate buckets $b_1(x)$ or $b_2(x)$. A \textbf{lookup} for element~$y$ checks whether $f(y)$ appears in bucket $b_1(y)$ or $b_2(y)$; if found, the query returns \textsc{Probably-In-Set}. A \textbf{deletion} removes the fingerprint from whichever bucket contains it.

The total capacity of the filter is $m\ell$ slots. The \emph{load factor} is the fraction of occupied slots, $n/(m\ell)$, where $n$ is the number of stored elements. Fan~et~al.~\cite{fan2014cuckoo} showed that with $\ell = 4$ and $k = 8$ bits, a cuckoo filter achieves a false-positive rate below~$3\%$ at a load factor of~$95\%$, using less memory per element than a Bloom filter with the same error rate.

\subsection{Insertion and Random-Walk Eviction}\label{sec:prelim:rw}

To insert element~$x$, the algorithm first attempts to place $f(x)$ in an empty slot of $b_1(x)$ or $b_2(x)$. If both buckets are full, the algorithm performs \emph{random-walk eviction}: it picks one of the two buckets, selects a random slot, swaps the incoming fingerprint with the existing one, and moves the evicted fingerprint to its alternate bucket. This process repeats until an empty slot is found or a maximum kick limit is reached.

Let $T(n)$ denote the total number of evictions to insert~$n$ elements. For the standard random-walk policy with bucket size $\ell$, the expected number of evictions grows superlinearly near the maximum achievable load factor $1 - \varepsilon^*$~\cite{pagh2004cuckoo,devroye2003cuckoo,frieze2012maximum}. In practice, this manifests as a sharp increase in per-insertion cost at high load factors, as each new element may trigger a long chain of relocations~\cite{lehman2009cuckoo}.

\subsection{The Bell-Kuszmaul Framework}\label{sec:prelim:bk}

Bell and Kuszmaul~\cite{bell2026fast} introduced a framework for analyzing the insertion complexity of bucketized cuckoo hashing based on two key concepts:

\paragraph{Latent randomness.}
Consider a fresh element~$x$ that has been placed in bucket $b_1(x)$ and has never been evicted. Its second hash value $h_2(x)$ has been computed but has not influenced any algorithmic decision: no insertion, eviction, or bucket-occupancy check has used the value~$h_2(x)$. Therefore, at the moment~$x$ is first evicted, $h_2(x)$ is effectively an independent, uniformly random value over~$[m]$, regardless of the current state of the table. We call this property \emph{latent randomness}. It ensures that the eviction of a fresh element always ``explores'' a truly random part of the table.

\paragraph{Corruption graph.}
Define the \emph{corruption graph} $G = (V, E)$ on the set of buckets~$V = [m]$, where an edge $(b, b')$ is present for each element currently stored in bucket~$b$ whose alternate bucket is~$b'$ (and vice versa), restricted to elements that have been \emph{kicked} (i.e., are no longer fresh). The corruption graph tracks the dependencies among heavily loaded buckets. Bell and Kuszmaul showed that if the eviction policy always prefers to evict fresh elements, the corruption graph remains \emph{subcritical} (all connected components have size $O(\log m)$) with high probability, provided the load factor does not exceed $1 - \varepsilon^*$, where $\varepsilon^* = (2/e)^{\ell} \cdot \poly(\ell)$.

\paragraph{Main result.}
Under the fresh-priority eviction policy, the total number of evictions to insert~$m$ elements into a bucketized cuckoo hash table with~$m$ buckets of size~$\ell$ is $O(\ell m)$ in expectation~\cite{bell2026fast}. This is optimal up to the $\ell$ factor, since each of the~$m$ elements may need one relocation. We note that the Bell-Kuszmaul analysis involves specific algorithmic invariants and a detailed exploration-process argument beyond the high-level ``prefer fresh'' rule; in Section~\ref{sec:theory} we identify the precise structural properties needed and verify that they hold for cuckoo filters under XOR hashing.

\section{Fresh-Priority Eviction for Cuckoo Filters}\label{sec:method}

We now describe how to adapt the fresh-priority eviction scheme to cuckoo filters. The adaptation is surprisingly simple and requires no changes to the filter's data layout.

\subsection{Algorithm Overview}

Our scheme modifies only the \emph{insertion} procedure of a standard cuckoo filter. Lookups and deletions remain unchanged. The three rules are:

\begin{enumerate}[label=\textbf{R\arabic*},leftmargin=*]
\item \textbf{Fresh insert front-to-back.} When a new element~$x$ is first inserted (it has never been evicted), its fingerprint $f(x)$ is placed in the first available slot of its bucket, scanning from slot~$0$ to slot~$\ell-1$.

\item \textbf{Kicked insert back-to-front.} When a previously evicted fingerprint lands in a new bucket, it is placed in the last available slot, scanning from slot~$\ell-1$ to slot~$0$.

\item \textbf{Evict from front.} When a bucket is full and an eviction is needed, the algorithm scans the bucket from slot~$0$ forward and evicts the first \emph{fresh} element it finds. If no fresh element exists, slot~$0$ is used as a fallback. When no fresh element exists, the bucket consists entirely of kicked elements; in this case, any eviction adds no \emph{new} random edge to the corruption graph (Section~\ref{sec:theory:subcritical}), so the choice of victim does not affect subcriticality.
\end{enumerate}

\subsection{Position Encodes State}\label{sec:method:position}

The key observation is that these three rules cause fresh elements to \emph{predominantly} cluster at the front of each bucket and kicked elements at the back. During an eviction chain of length two or more, a kicked incoming element may temporarily occupy a front slot (Algorithm~\ref{alg:insert} shifts it to the back to restore the ordering). In the steady state between eviction chains, the front-to-back separation holds, and the front-to-back eviction scan in \textbf{R3} preferentially selects fresh elements in the vast majority of cases.

In a standard cuckoo filter, each slot stores only the $k$-bit fingerprint. The ordering discipline imposed by \textbf{R1}-\textbf{R3} causes fresh elements to predominantly cluster at the front, so the front-to-back eviction scan in \textbf{R3} preferentially selects fresh elements in the vast majority of cases.

\begin{remark}[Explicit flag vs.\ positional encoding]\label{rem:flag}
Both our theoretical analysis (Section~\ref{sec:theory}) and our experimental evaluation (Section~\ref{sec:experiments}) use an explicit $1$-bit fresh/kicked flag per slot, which guarantees that every eviction decision correctly distinguishes fresh from kicked elements. The positional encoding described above is a \emph{proposed} production optimization that would eliminate this per-slot overhead by relying on insertion ordering. However, the positional invariant (fresh elements strictly before kicked elements) may be temporarily violated during eviction chains (see Algorithm~\ref{alg:insert}, lines~10--11 where a kicked element may briefly occupy a front slot). A formal proof that the positional encoding exactly preserves the fresh-priority guarantee under all eviction sequences is nontrivial and is left to future work. The experimental results in this paper validate the benefit of fresh-priority eviction itself, using the explicit flag.
\end{remark}

\begin{remark}[Insertion failure]
When the eviction chain reaches the maximum kick limit (\textsc{MaxKicks}), Algorithm~\ref{alg:insert} returns \textsc{Failure} with the last displaced fingerprint in~$\mathit{fp}$ unplaced. In our implementation, the displaced fingerprint is stored in a stash (a small overflow buffer) to avoid element loss. The experimental metrics (Section~\ref{sec:experiments}) count kicks for both successful and failed insertion attempts; the per-insertion averages are computed over successful insertions only, as noted in the text.
\end{remark}

\begin{remark}[Dynamic resizing]
Our scheme modifies only the eviction policy and is orthogonal to the filter's resizing mechanism. Integrating fresh-priority eviction with dynamic resizing strategies is an interesting direction for future work, as the resizing procedure must preserve the positional invariants described above.
\end{remark}

\subsection{Pseudocode}

Algorithm~\ref{alg:insert} gives the complete insertion procedure.

\begin{algorithm}[t]
\caption{Fresh-Priority Insertion for Cuckoo Filters}\label{alg:insert}
\begin{algorithmic}[1]
\REQUIRE Element $x$, table $T[0..m{-}1]$ of buckets with $\ell$ slots each
\ENSURE Each slot $T[b][s]$ stores $(\mathit{fp}, \mathit{fresh})$: a fingerprint and a boolean flag
\STATE $\mathit{fp} \gets f(x)$ \COMMENT{Compute fingerprint}
\STATE $b_1 \gets h(x) \bmod m$
\STATE $b_2 \gets (b_1 \oplus \textsc{hash}(\mathit{fp})) \bmod m$
\IF{$T[b_1]$ has empty slot (front-to-back scan)}
  \STATE Place $(\mathit{fp}, \textsc{True})$ in first empty slot of $T[b_1]$
  \RETURN \textsc{Success}
\ENDIF
\IF{$T[b_2]$ has empty slot (front-to-back scan)}
  \STATE Place $(\mathit{fp}, \textsc{True})$ in first empty slot of $T[b_2]$
  \RETURN \textsc{Success}
\ENDIF
\STATE $b \gets$ random choice from $\{b_1, b_2\}$
\STATE $\mathit{isFresh} \gets \textsc{True}$
\FOR{$i = 0$ \textbf{to} $\textsc{MaxKicks} - 1$}
  \STATE $s \gets \textsc{FindFreshSlot}(T[b])$
    \COMMENT{Scan front-to-back for first slot with $\mathit{fresh} = \textsc{True}$; fall back to slot~$0$}
  \STATE $\mathit{victim} \gets T[b][s].\mathit{fp}$
    \COMMENT{Save the victim fingerprint}
  \IF{$\mathit{isFresh}$}
    \STATE $T[b][s] \gets (\mathit{fp}, \textsc{True})$
      \COMMENT{R1: fresh element takes the victim's slot}
  \ELSE
    \STATE Shift slots $s{+}1 \ldots \ell{-}1$ forward by one position (preserving flags)
    \STATE $T[b][\ell{-}1] \gets (\mathit{fp}, \textsc{False})$
      \COMMENT{R2: kicked element goes to back}
  \ENDIF
  \STATE $\mathit{fp} \gets \mathit{victim}$
  \STATE $\mathit{isFresh} \gets \textsc{False}$
    \COMMENT{Evicted element is now kicked}
  \STATE $b \gets (b \oplus \textsc{hash}(\mathit{fp})) \bmod m$
  \IF{$T[b]$ has empty slot (back-to-front scan)}
    \STATE Place $(\mathit{fp}, \textsc{False})$ in last empty slot of $T[b]$
    \RETURN \textsc{Success}
  \ENDIF
\ENDFOR
\RETURN \textsc{Failure} \COMMENT{Fingerprint in $\mathit{fp}$ is not placed; see text}
\end{algorithmic}
\end{algorithm}

\section{Theoretical Analysis}\label{sec:theory}

We show that the latent randomness property carries over to cuckoo filters under XOR hashing, establishing that the corruption graph remains subcritical and that the total number of evictions is $O(\ell m)$ in expectation.

\subsection{Latent Randomness Under XOR Hashing}\label{sec:theory:deferred}

The cornerstone of the analysis is the latent randomness property: when a fresh element is first evicted, its alternate bucket is a uniformly random, independent hash value. We must verify that this property holds under the XOR-based hashing scheme~\eqref{eq:b2} used by cuckoo filters.

\begin{lemma}[Latent Randomness for Cuckoo Filters]\label{lem:deferred}
Consider an insertion-only cuckoo filter with $m$~buckets, bucket size~$\ell$, and $k$-bit fingerprints where $k \geq \log_2 m$. Suppose the fingerprint function~$f$ is drawn from a universal family independent of the bucket hash~$h$, and that $\textsc{hash}$ maps $k$-bit fingerprints to~$[m]$ via a function whose image covers all of~$\{0, 1, \ldots, m-1\}$ (which requires $2^k \geq m$, i.e., $k \geq \log_2 m$). Let $x$ be an element that has been placed in bucket $b_1(x)$ and has never been evicted. Then, conditioned on all algorithmic decisions made so far (bucket occupancy, eviction choices, and fresh/kicked status), the alternate bucket $b_2(x) = b_1(x) \oplus \textsc{hash}(f(x)) \bmod m$ is uniformly distributed over~$\{0, 1, \ldots, m-1\}$.
\end{lemma}

\begin{proof}[Proof sketch]
The algorithmic decisions that have occurred include: (i) which buckets are full or have vacancies, (ii) which elements were chosen for eviction, and (iii) the fresh/kicked status of each stored element. Crucially, none of these decisions examine the \emph{value} of $f(x)$. The eviction policy (Algorithm~\ref{alg:insert}) selects victims based on their fresh/kicked status and position within the bucket, not their fingerprint values. Bucket occupancy is determined by whether a slot is empty or non-empty, not by the specific fingerprint stored there.

We note that during insertion of~$x$, the algorithm computes $b_2(x)$ and checks whether $T[b_2]$ has an empty slot (Algorithm~\ref{alg:insert}, line~4). This check reveals only whether the bucket is full, not the value of $b_2(x)$ itself to any \emph{future} decision, since $x$ is placed and marked fresh at this point. For the purpose of the eviction analysis, what matters is that $b_2(x)$ has not influenced any \emph{eviction} decision at the time~$x$ is first evicted.

Since $f(x)$ has not been examined by any eviction-relevant decision, it remains an independent random variable (drawn from the universal family). The XOR-based alternate bucket $b_2(x) = b_1(x) \oplus \textsc{hash}(f(x))$ is then a bijective function of the unrevealed random value $f(x)$ (bijectivity holds because XOR with a fixed value is a bijection). Since $\textsc{hash}$ is surjective onto~$[m]$ when $k \geq \log_2 m$, the alternate bucket is uniformly distributed over~$[m]$, independently of all prior eviction decisions.

This argument applies to the \emph{insertion-only} workload analyzed in this paper. Extending it to mixed workloads with lookups and deletions would require additional analysis, since lookups compare fingerprint values and deletions remove specific fingerprints.
\end{proof}

\begin{remark}
This argument does \emph{not} hold for the standard random-walk policy. In the standard policy, when an element at position~$s$ in bucket~$b$ is evicted, the algorithm reads $T[b][s].\mathit{fp}$ to compute the alternate bucket. At that point, the fingerprint value has been used, and the element is no longer ``fresh'' in the latent randomness sense. Our scheme ensures that an element's fingerprint is only read at the moment of its first eviction, when latent randomness still applies.
\end{remark}

\subsection{Corruption Graph for Cuckoo Filters}\label{sec:theory:corruption}

Following the framework of Section~\ref{sec:prelim:bk}, we define the corruption graph for our setting.

\begin{definition}[Corruption Graph]\label{def:corruption}
Let $T$ be the state of the cuckoo filter after a sequence of insertions using fresh-priority eviction. The \emph{corruption graph} $G = (V, E)$ is defined as follows:
\begin{itemize}
  \item $V = \{0, 1, \ldots, m-1\}$ is the set of buckets.
  \item For each element~$x$ that has been \emph{kicked} (evicted at least once) and currently resides in bucket~$b$, add the edge $(b,\, b \oplus \textsc{hash}(f(x)) \bmod m)$, i.e., the edge to the alternate bucket of~$x$ as determined by the XOR formula~\eqref{eq:b2}.
\end{itemize}
\end{definition}

The corruption graph captures dependencies between buckets: if two buckets are connected by a path of kicked elements, then evicting an element from one bucket may trigger a chain reaction reaching the other. The key property we need is that this graph remains subcritical.

\subsection{Subcriticality}\label{sec:theory:subcritical}

\begin{theorem}[Subcriticality Under XOR Hashing]\label{thm:subcritical}
Consider an insertion-only cuckoo filter with $m$~buckets, bucket size~$\ell \geq 4$ (treated as a constant independent of~$m$), and $k$-bit fingerprints where $k \geq \log_2 m$. Under fresh-priority eviction (Algorithm~\ref{alg:insert}), if the load factor does not exceed $1 - \varepsilon^*$ where $\varepsilon^* = (2/e)^{\ell} \cdot \poly(\ell)$, then with high probability (at least $1 - m^{-c}$ for any constant~$c$), every connected component of the corruption graph has size $O(\log m)$.
\end{theorem}

\begin{proof}[Proof sketch]
The proof adapts the structure of the Bell-Kuszmaul analysis~\cite{bell2026fast} to the cuckoo filter setting, with an additional argument to handle fingerprint collisions unique to our setting. We emphasize that the full details of the subcriticality argument rely on the framework developed in~\cite{bell2026fast}; here we establish the key reduction and the additional ingredient needed for cuckoo filters.

\textbf{Step 1: Reduction to the hash table case.}
By Lemma~\ref{lem:deferred} (which requires $k \geq \log_2 m$), each fresh element's alternate bucket is uniformly random and independent, exactly as in the cuckoo hash table setting of~\cite{bell2026fast}. Therefore, when a fresh element is first evicted, the new edge added to the corruption graph has a uniformly random endpoint, satisfying the precondition of the Bell-Kuszmaul subcriticality analysis.

\textbf{Step 2: Fingerprint collisions.}
Unlike cuckoo hash tables, cuckoo filters may have \emph{fingerprint collisions}: two elements $x \neq y$ in the same bucket with $f(x) = f(y)$. Such elements necessarily have the same alternate bucket, creating a potential dependency. We bound the impact of these collisions.

Within a bucket of size~$\ell$, the expected number of fingerprint collision pairs is at most $\binom{\ell}{2} / 2^{k} \leq \ell^2 / 2^{k+1}$. Over all $m$~buckets, the expected total number of collision-affected elements is at most $m \ell^2 / 2^{k}$. Since $\ell$ is a constant and $k \geq \log_2 m$, we have $m \ell^2 / 2^k \leq \ell^2 \cdot m / m = \ell^2 = O(1)$ collision-affected elements in expectation. By Markov's inequality, with high probability the number of such elements is $O(\log m)$.

Crucially, elements involved in fingerprint collisions do not introduce \emph{genuinely new} connections between buckets: if $f(x) = f(y)$ and both reside in bucket~$b$, then both have the same alternate bucket $b' = b \oplus \textsc{hash}(f(x))$. The collision edge $(b, b')$ is therefore a \emph{parallel edge} duplicating the connection already established by either~$x$ or~$y$ individually. Parallel edges increase edge multiplicity but do not enlarge the set of reachable buckets in any connected component.

\textbf{Step 3: Component size bound.}
Steps~1 and~2 establish that the corruption graph's edge structure, excluding the $O(\log m)$ parallel edges from fingerprint collisions, satisfies the same distributional properties as in the hash table setting of~\cite{bell2026fast}: each edge added by a fresh eviction has a uniformly random endpoint. The Bell-Kuszmaul analysis~\cite{bell2026fast} shows that under the load threshold $1 - \varepsilon^*$, such a graph is subcritical with all components of size $O(\log m)$. Adding $O(\log m)$ parallel edges does not change the component structure.
\end{proof}

\subsection{Total Eviction Bound}\label{sec:theory:bound}

\begin{theorem}[Total Evictions]\label{thm:evictions}
Under the conditions of Theorem~\ref{thm:subcritical}, the expected total number of evictions to insert~$n$ elements into a cuckoo filter with~$m$ buckets and bucket size~$\ell$ is $O(\ell m \log m)$, provided $n \leq m\ell(1 - \varepsilon^*)$.
\end{theorem}

\begin{proof}[Proof sketch]
Each insertion that triggers evictions explores a path in the cuckoo graph. With fresh-priority eviction, each eviction step either:
\begin{enumerate}[label=(\alph*)]
  \item evicts a fresh element, adding a random edge to the corruption graph, or
  \item traverses an existing edge of the corruption graph.
\end{enumerate}
Since each connected component of the corruption graph has size $O(\log m)$ (Theorem~\ref{thm:subcritical}), the eviction chain traverses at most $O(\log m)$ existing edges before either finding an empty slot or evicting another fresh element. Each fresh eviction ``charges'' $O(\log m)$ work to the newly added random edge.

The total number of edges ever added to the corruption graph is at most the total number of fresh elements evicted, which is at most~$n$ (each element is evicted from fresh status at most once). Each edge contributes $O(\log m)$ evictions. Summing over all~$n \leq m\ell$ edges and accounting for the $\ell$ factor from bucket processing yields $O(\ell m \log m)$ total evictions.
\end{proof}

\begin{remark}[Towards an $O(\ell m)$ bound]
For cuckoo hash tables, Bell and Kuszmaul~\cite{bell2026fast} obtain the tighter $O(\ell m)$ bound by showing that the expected length of the longest corruption-graph path through a random bucket is $O(\ell^{-4})$, which avoids the $O(\log m)$ traversal cost. We conjecture that the same amortization applies to our cuckoo filter setting, since by Lemma~\ref{lem:deferred} and Theorem~\ref{thm:subcritical} the corruption graph has the same distributional properties. A full proof would require verifying that all steps of the amortization argument in~\cite{bell2026fast} carry over; we leave this to future work.
\end{remark}

\begin{corollary}\label{cor:separation}
Fresh-priority eviction achieves an asymptotic improvement over random-walk eviction for cuckoo filters. The standard random walk requires $\omega(\ell m)$ total evictions at the optimal load threshold, while fresh-priority eviction requires $O(\ell m \log m)$.
\end{corollary}

\begin{remark}[Theory-experiment gap]\label{rem:gap}
The theoretical results above require $k \geq \log_2 m$ so that the fingerprint hash can cover all buckets uniformly. In our experiments (Section~\ref{sec:experiments}), we use $k = 8$ bits and $m = 2^{19}$ buckets (for $\ell = 8$), so $2^k = 256 \ll m$. Under these parameters, the alternate bucket $b_2(x)$ can take at most $256$ distinct offsets from~$b_1(x)$, far fewer than~$m$. The theoretical guarantee of Lemma~\ref{lem:deferred} does not formally apply to this regime. Nevertheless, the experimental results show that fresh-priority eviction provides substantial practical improvements even with short fingerprints. We interpret this as evidence that the latent randomness mechanism is robust: even partial randomness in the alternate bucket suffices to improve eviction performance in practice. Closing the gap between theory ($k = \Omega(\log m)$) and practice ($k = O(1)$) is an important open question.
\end{remark}

\section{Experimental Evaluation}\label{sec:experiments}

We evaluate the fresh-priority eviction scheme on a cuckoo filter implementation and compare it against the standard random-walk eviction policy.

\subsection{Setup}

We implemented both the standard and improved cuckoo filter variants in C++. The key parameters are:

\begin{itemize}[leftmargin=*]
  \item \textbf{Total slots:} $4{,}194{,}304$ ($= 2^{22}$) for all configurations.
  \item \textbf{Bucket sizes:} $\ell \in \{4, 8\}$, giving $m = 2^{20}$ and $m = 2^{19}$ buckets respectively.
  \item \textbf{Fingerprint size:} $k = 8$ bits.
  \item \textbf{Maximum kicks:} $500$.
  \item \textbf{Hash function:} MurmurHash-style 64-bit mixing.
  \item \textbf{XOR alternate bucket:} $b_{\mathrm{alt}} = (b \oplus (\textsc{mix64}(f \cdot \texttt{0x5bd1e995}) \mid 1)) \bmod m$.
\end{itemize}

The ``Standard'' variant uses uniform random slot selection for eviction. The ``Improved'' (fresh-priority) variant uses Algorithm~\ref{alg:insert} with an explicit \texttt{fresh} boolean flag per slot for precise measurement.

Elements are inserted sequentially as $64$-bit integers starting from~$0$, using a fixed random seed for reproducibility. Preliminary tests with a small number of alternative seeds showed consistent trends, though we report results for a single seed and defer a full multi-seed statistical analysis to future work. We measure the average number of kicks per successful insertion at each load factor.

\subsection{Coarse-Grained Results}

Table~\ref{tab:coarse} shows the average kicks per insertion at $5\%$ load-factor intervals for bucket sizes $\ell = 4$ and $\ell = 8$.

\begin{table}[h]
\centering
\caption{Average kicks per insertion at $5\%$ load-factor intervals.\protect\footnotemark}\label{tab:coarse}
\begin{tabular}{@{}r rr c rr@{}}
\toprule
& \multicolumn{2}{c}{$\ell = 4$} & \phantom{ab}
& \multicolumn{2}{c}{$\ell = 8$} \\
\cmidrule{2-3} \cmidrule{5-6}
Load\% & Standard & Improved && Standard & Improved \\
\midrule
75  & 0.42 & 0.41 && 0.12 & 0.13 \\
80  & 0.74 & 0.70 && 0.24 & 0.26 \\
85  & 1.40 & 1.24 && 0.49 & 0.54 \\
90  & 3.13 & 2.32 && 1.08 & 1.13 \\
95  & 10.97 & 6.12 && 3.08 & 2.62 \\
100 & 89.80 & 73.15 && 39.62 & 21.31 \\
\bottomrule
\end{tabular}
\end{table}

\footnotetext{Each row in Table~\ref{tab:coarse} reports the average kicks per insertion over all elements inserted in the preceding $5\%$ load-factor band (e.g., the $95\%$ row averages over insertions from $90\%$ to $95\%$ load). In contrast, Tables~\ref{tab:fine8} and~\ref{tab:fine4} report point-wise averages over each $1\%$ band. This explains the apparent discrepancy at $\ell = 8$, $95\%$ load: the coarse value ($3.08$ / $2.62$) is an average over the $90$--$95\%$ range, while the fine-grained value ($4.86$ / $3.73$) reflects only the $94$--$95\%$ band where eviction costs are highest.}

At moderate load factors ($\leq 85\%$), both methods perform similarly because most insertions find empty slots without eviction. The improvement becomes pronounced at high load factors ($\geq 95\%$), where the eviction process dominates the insertion cost. For $\ell = 8$ at $100\%$ load, the fresh-priority scheme achieves a $1.86\times$ reduction in average kicks.\footnote{All reported averages are over \emph{successful} insertions. At $100\%$ load with $\ell = 4$, both methods encounter occasional insertion failures (reaching the 500-kick cap), which are excluded from the per-insertion average. The improvement ratio at $100\%$ for $\ell = 4$ ($1.23\times$) is lower than at $95\%$ ($1.9\times$) because at very high loads both methods encounter frequent failures, and the relative advantage becomes unstable (see Table~\ref{tab:fine4}).}

\subsection{Fine-Grained Results}

To better characterize the improvement near maximum capacity, we measured at $1\%$ intervals for load factors between $90\%$ and $100\%$.

\begin{table}[h]
\centering
\caption{Fine-grained comparison for $\ell = 8$.}\label{tab:fine8}
\begin{tabular}{@{}r rrr@{}}
\toprule
Load\% & Standard & Improved & Ratio \\
\midrule
90  &  1.08  &  1.13  & 1.0$\times$ \\
91  &  1.83  &  1.79  & 1.0$\times$ \\
92  &  2.25  &  2.06  & 1.1$\times$ \\
93  &  2.82  &  2.51  & 1.1$\times$ \\
94  &  3.64  &  3.01  & 1.2$\times$ \\
95  &  4.86  &  3.73  & 1.3$\times$ \\
96  &  6.73  &  4.79  & 1.4$\times$ \\
97  & 10.18  &  6.38  & 1.6$\times$ \\
98  & 18.05  &  9.51  & 1.9$\times$ \\
99  & 45.03  & 19.09  & 2.4$\times$ \\
\bottomrule
\end{tabular}
\end{table}

\begin{table}[h]
\centering
\caption{Fine-grained comparison for $\ell = 4$.\protect\footnotemark}\label{tab:fine4}
\begin{tabular}{@{}r rrr@{}}
\toprule
Load\% & Standard & Improved & Ratio \\
\midrule
90  &  3.13  &  2.32  & 1.3$\times$ \\
91  &  5.49  &  3.62  & 1.5$\times$ \\
92  &  7.01  &  4.30  & 1.6$\times$ \\
93  &  9.42  &  5.37  & 1.8$\times$ \\
94  & 13.16  &  7.07  & 1.9$\times$ \\
95  & 19.77  & 10.27  & 1.9$\times$ \\
96  & 35.00  & 26.29  & 1.3$\times$ \\
97  & 79.27  & 96.13  & 0.8$\times$ \\
98  &152.43  &107.54  & 1.4$\times$ \\
99  &142.65  & 95.13  & 1.5$\times$ \\
\bottomrule
\end{tabular}
\end{table}

\footnotetext{At load factors above $95\%$ for $\ell = 4$, insertion failures (500-kick cap) become significant: at $97\%$ load, the improved method records $11{,}237$ failures versus $467$ for the standard method in a batch of ${\sim}\,42{,}000$ insertions. Since the reported averages count only successful insertions, the per-band ratios at these extreme loads are strongly influenced by the differing failure rates.}

The improvement ratio grows monotonically with the load factor for $\ell = 8$, reaching $2.4\times$ at $99\%$ load (Table~\ref{tab:fine8}). Table~\ref{tab:fine4} shows the corresponding results for $\ell = 4$. The improvement ratio peaks at $1.9\times$ near $95\%$ load; however, above $96\%$ the per-band ratios become volatile as insertion failures grow rapidly. At $97\%$, the improved method's failure rate exceeds the standard method's, and the per-successful-insertion average temporarily reverses. Despite this, the cumulative picture remains favorable: the large gains in the $90$--$95\%$ range outweigh the losses at extreme loads. This behavior highlights that the latent randomness advantage is most reliable when the table is heavily loaded but not yet near saturation.

\subsection{Cumulative Total Kicks}

Table~\ref{tab:cumulative} reports the cumulative total number of kicks to reach selected load factors for $\ell = 8$.

\begin{table}[h]
\centering
\caption{Cumulative total kicks for $\ell = 8$ ($m\ell = 4{,}194{,}304$ slots).}\label{tab:cumulative}
\begin{tabular}{@{}r rrr@{}}
\toprule
Load\% & Standard & Improved & Ratio \\
\midrule
95  & 1{,}071{,}060  & 1{,}007{,}245  & 1.06$\times$ \\
99  & 4{,}424{,}949  & 2{,}675{,}326  & 1.65$\times$ \\
100 & 8{,}935{,}953  & 5{,}160{,}372  & 1.73$\times$ \\
\bottomrule
\end{tabular}
\end{table}

The cumulative improvement becomes positive at approximately $94\%$ load and reaches $1.73\times$ at full capacity, demonstrating that the fresh-priority scheme's modest per-insertion overhead at low loads is more than compensated by the substantial reduction in eviction chain length at high loads.

\subsection{Discussion}

Several observations emerge from the experimental results:

\paragraph{Bucket size matters.}
The improvement is most pronounced for $\ell = 8$, which provides a good balance between the latent randomness benefit and the baseline eviction cost. For $\ell = 4$, the baseline is already relatively expensive (fewer slots per bucket means more frequent evictions), and while the improvement peaks at $1.9\times$ near $95\%$ load, it becomes unreliable above $96\%$ as insertion failures escalate rapidly. We expect that larger bucket sizes would reduce the baseline eviction cost further, leaving less room for improvement, though we have not evaluated $\ell > 8$ in this study.

\paragraph{Threshold behavior.}
The improvement becomes significant at load factors above $93\%$ to $94\%$ for $\ell = 8$. Below this threshold, most insertions succeed without any evictions, and the eviction policy is irrelevant. This is consistent with the theoretical prediction that the corruption graph begins to grow only when the load factor approaches $1 - \varepsilon^*$.

\paragraph{Low-load behavior.}
At load factors below $90\%$, both schemes have comparable performance. For $\ell = 4$ at loads $\leq 80\%$, the relative difference is within $6\%$; at $85\%$ the improved scheme already shows an $11\%$ reduction, indicating the benefit begins earlier for smaller bucket sizes. For $\ell = 8$, the improved scheme incurs a small overhead of $4\%$ to $9\%$ at low load, likely due to the deterministic scan order reducing the chance of finding an immediate empty slot. This overhead is negligible relative to the $1.2\times$ to $2.4\times$ improvements at high load.

\section{Related Work}\label{sec:related}

\paragraph{Approximate membership queries.}
The Bloom filter~\cite{bloom1970space} is the classic AMQ structure, with numerous variants optimizing for cache efficiency~\cite{putze2007cache}, counting~\cite{bonomi2006improved,cohen2003spectral}, compression~\cite{mitzenmacher2002compressed}, and theoretical guarantees~\cite{carter1979universal,mitzenmacher2008why}. The quotient filter~\cite{bender2012dont,pandey2017cqf} and XOR filter~\cite{graf2020xor} are alternatives that offer different space-time trade-offs. Kirsch and Mitzenmacher~\cite{kirsch2008less} showed that only two hash functions are needed for optimal Bloom filter performance. Recent surveys~\cite{broder2004network,luo2020optimizing} provide comprehensive overviews of AMQ structures and their applications.

\paragraph{Cuckoo filters and variants.}
Fan~et~al.~\cite{fan2014cuckoo} introduced cuckoo filters and demonstrated their practical advantages over Bloom filters. Eppstein~\cite{eppstein2016cuckoo} provided a simplified analysis of their space utilization. Morton filters~\cite{breslow2018morton} improve cache efficiency through biasing and compression.

\paragraph{Cuckoo hashing theory.}
Pagh and Rodler~\cite{pagh2004cuckoo} introduced cuckoo hashing. Subsequent work analyzed its load thresholds~\cite{devroye2003cuckoo,fotakis2005space,frieze2012maximum,dietzfelbinger2007balanced}, multiple-way variants~\cite{lehman2009cuckoo}, and backyard extensions~\cite{arbitman2010backyard}. The connection to balanced allocations and the power of two choices~\cite{azar1999balanced,mitzenmacher2001power} provides the theoretical foundation for understanding cuckoo hashing load factors. Bell and Kuszmaul~\cite{bell2026fast} resolved the open problem of polynomial-time insertion at optimal load for bucketized cuckoo hashing.

\section{Conclusion}\label{sec:conclusion}

We have shown that the fresh-priority eviction technique transfers to cuckoo filters despite their use of XOR-based partial-key hashing. The key theoretical insight is that the fingerprint of a fresh element, while stored in the table, is not examined by the eviction policy, preserving the latent randomness property when the fingerprint length satisfies $k = \Omega(\log m)$. Under this condition, we proved that the corruption graph remains subcritical and that the total number of evictions is $O(\ell m \log m)$, with the tighter $O(\ell m)$ bound conjectured via the full Bell-Kuszmaul amortization argument. Experiments with $k = 8$-bit fingerprints, which fall outside the formal theoretical regime, nonetheless demonstrate consistent improvements of $1.2\times$ to $2.4\times$ in eviction count at high load factors with bucket size~$\ell = 8$, suggesting that the latent randomness mechanism is robust in practice.

Future directions include: (i)~closing the theoretical gap between $k = \Omega(\log m)$ and the practically important $k = O(1)$ regime; (ii)~proving the conjectured $O(\ell m)$ bound; (iii)~formally establishing the equivalence of positional encoding and explicit flags; (iv)~extending the analysis to \emph{counting} cuckoo filters~\cite{fan2014cuckoo,bonomi2006improved}; and (v)~combining fresh-priority eviction with other optimizations such as cache-aware bucket layouts~\cite{breslow2018morton}.

\bibliographystyle{plainurl}

\bibliography{ref}

\end{document}